\documentstyle[prl,aps]{revtex}

\begin{document}
\title{{\bf Improved scheme for generation of vibrational trio coherent states of a
trapped ion}}
\author{Hyo Seok Yi}
\address{Department of Physics, Korea Advanced Institute of Science and Technology,\\
373-1 Guseong-dong, Yuseong-gu, Daejeon 305-701, Republic of Korea}
\author{Ba An Nguyen\thanks{%
Corresponding author. Email: nbaan@kias.re.kr} and Jaewan Kim}
\address{School of Computational Sciences, Korea Institute for Advanced Study, \\
207-43 Cheongryangni 2-dong, Dongdaemun-gu, Seoul 130-722, Republic of Korea}
\maketitle
\date{}

\begin{abstract}
We improve a previously proposed scheme (Phys. Rev. A 66 (2002) 065401) for
generating vibrational trio coherent states of a trapped ion. The improved
version is shown to gain a double advantage: (i) it uses only five, instead
of eight, lasers and (ii) the generation process can be made remarkably
faster.

\end{abstract}


Following the pair coherent state (PCS) \cite{pcs}, which has proved very
important in quantum optics \cite{qo}, atom-field dynamics \cite{afd},
quantum-mechanics-versus-local-realism test \cite{qmt}, continuous-variable
quantum nonlocality \cite{cvqn} and quantum information \cite{qi}, the
so-called trio coherent state (TCS) \cite{tcs} and their cat-type
superpositions \cite{cattrio} have also been introduced recently. This novel
family of states has been shown inherently nonclassical exhibiting new types
of multimode antibunching, higher-order squeezing and violation of
Cauchy-Schwarz inequalities.{\it \ }Although the recognized kinds of
nonclassical states have been numerous to date (see, {\it e.g}. \cite{r4d}),
it is not yet justified that all of them acquire actual applications. It is
also not excluded that ``really needed'' states remain undiscovered or they
are among the existing ones with their necessary properties unnoticeable. In
that sense, any further detailed study of a known state or introduction of a
new state would be perhaps equally welcome. As for the TCS, because of their
three-mode nature, they would play a significant role in phenomena in which
the PCS could not. For example, with the TCS one might predict that
``one-event'' continuous-variable local realism violating experiments could
be found in a way more or less similar to those devised by
Greenberger-Horne-Zeilinger \cite{ghz} or/and by Hardy \cite{hardy} for
observables with discrete spectra. However, before any expected experiment
can be done, a primary question arises: how can one produce the TCS in
practice ? This question was answered in \cite{tcs} for electromagnetic
fields interacting with an atom when there is strong competition between
trio parametric conversion and trio absorption. Here we shall concern with
generation of TCS in the center-of-mass motion of a trapped ion. The
motivation is that advanced laser-cooling techniques applied to single
trapped ions allow to cool them down near their zero-point vibrational
energy (see, {\it e.g., }Ref. \cite{lc}). In that quantized regime, the ion
vibrational modes can be looked upon as ideal bosons which are almost
uninfluenced by the environment resulting in negligible decoherences. This
fact greatly favors potential implementations such as designing quantum
logic gates in a quantum computer \cite{qc} (in particular, the
controlled-NOT gate has been realized in two $^{40}Ca^{+}$ ions in a Paul
trap individually radiated by focused lasers \cite{cnot}). Laser-driven
trapped ions placed inside optical cavities can also entangle the cavity
mode simultaneously with the internal and external degrees of freedom of the
ion \cite{ent}. Previously, the vibrational TCS in a 3D trap was shown to be
generated by a scheme using eight lasers \cite{tam} (compare with schemes
for the vibrational PCS in a 2D trap \cite{2D}). In this Letter we shall
improve the scheme proposed in \cite{tam} by reducing the number of used
lasers from eight to five. Moreover, we shall carry out detailed simulation
to reveal that the time needed to generate the vibrational TCS can be made
shorter in the improved version than in the original one \cite{tam}.

The vibrational TCS, denoted by $\left| \xi ,p,q\right\rangle $ with $\xi
=r\exp (i\varphi )$ and $p,q$ integers, is a three-mode entangled continuous
superposition of coherent states of the form 
\begin{eqnarray}
\left| \xi ,p,q\right\rangle &=&\frac{N(p,q,r)\exp (3r^{2/3}/2)}{\xi
^{(p+q)/3}}  \nonumber \\
&&\times \int_{0}^{2\pi }\frac{d\theta }{2\pi }\int_{0}^{2\pi }\frac{d\theta
^{\prime }}{2\pi }\exp [-i(q\theta +p\theta ^{\prime })]  \nonumber \\
&&\times \left| \xi ^{1/3}\exp (i\theta )\right) _{x}\left| \xi ^{1/3}\exp
(i\theta ^{\prime })\right) _{y}\left| \xi ^{1/3}\exp [-i(\theta +\theta
^{\prime })]\right) _{z}  \label{cs}
\end{eqnarray}
where $N(p,q,r)$ defined by 
\begin{equation}
N(p,q,r)^{-2}=\sum_{n=0}^{\infty }\frac{r^{2n}}{(n+q)!(n+p)!n!}  \label{N}
\end{equation}
is the normalization coefficient and $\left| ...\right) _{x(y,z)}$ stand for
coherent states in the vibrational mode along the $x$ $(y,z)$ axis with the
annihilation boson operators $\widehat{a}_{x(y,z)}.$ Equivalently, the state 
$\left| \xi ,p,q\right\rangle $ can also be represented in terms of
correlated boson number trios as 
\begin{equation}
\left| \xi ,p,q\right\rangle =\sum_{n=0}^{\infty }C_{n}(\xi ,p,q)\left|
n+q\right\rangle _{x}\left| n+p\right\rangle _{y}\left| n\right\rangle _{z}
\label{fs}
\end{equation}
where 
\begin{equation}
C_{n}(\xi ,p,q)=\frac{N(p,q,r)\xi ^{n}}{\sqrt{(n+q)!(n+p)!n!}}  \label{C}
\end{equation}
and $\left| ...\right\rangle _{x(y,z)}$ stand for Fock states in the $x$ $%
(y,z)$ axis mode. Both the representations (\ref{cs}) and (\ref{fs}) come
from the fact that the vibrational TCS is defined as the joint eigenstate of
the operators $\widehat{a}_{x}\widehat{a}_{y}\widehat{a}_{z},$ $\widehat{P}=%
\widehat{a}_{y}^{+}\widehat{a}_{y}-\widehat{a}_{z}^{+}\widehat{a}_{z}$ and $%
\widehat{Q}=\widehat{a}_{x}^{+}\widehat{a}_{x}-\widehat{a}_{z}^{+}\widehat{a}%
_{z},$ {\it i.e.} 
\begin{equation}
\widehat{a}_{x}\widehat{a}_{y}\widehat{a}_{z}\left| \xi ,p,q\right\rangle
=\xi \left| \xi ,p,q\right\rangle ,  \label{e1}
\end{equation}
\begin{equation}
\widehat{P}\left| \xi ,p,q\right\rangle =p\left| \xi ,p,q\right\rangle ,
\label{e2}
\end{equation}
\begin{equation}
\widehat{Q}\left| \xi ,p,q\right\rangle =q\left| \xi ,p,q\right\rangle .
\label{e3}
\end{equation}

Generally, the system of a two-level ($\left| g\right\rangle $ and $\left|
e\right\rangle :$ ground and excited state) ion localized in a small region
by a 3D isotropic harmonic potential and radiated by $L$ external lasers is
described by the Hamiltonian $(\hbar =1)$%
\begin{equation}
H=H_{0}+H_{int},  \label{H}
\end{equation}
\begin{equation}
H_{0}=\frac{1}{2}\Delta \sigma _{z}+\sum_{j=x,y,z}\nu \left( \widehat{a}%
_{j}^{+}\widehat{a}_{j}+\frac{1}{2}\right) ,  \label{H0}
\end{equation}
\begin{equation}
H_{int}=\sum_{l=1}^{L}\left[ \Omega _{l}\exp [i(\omega _{l}t+\phi
_{l})]g_{l}\sigma _{-}+\text{H.c.}\right]  \label{Hint}
\end{equation}
with $\Delta $ the energy gap between the ion's two levels, $\sigma
_{-}=\left| g\right\rangle \left\langle e\right| =\sigma _{+}^{+},$ $\sigma
_{z}=\left| e\right\rangle \left\langle e\right| -\left| g\right\rangle
\left\langle g\right| ,$ $\nu $ the energy of a quantum of the vibrational
motion, $\Omega _{l}$ the Rabi frequencies, $\omega _{l}$ $(\phi _{l})$ the
laser frequencies (phases) and $g_{l}$ the laser spatial profile. For
travelling waves $g_{l}=\exp [-i{\bf k}_{l}{\bf R}_{l}]$ with ${\bf R}_{l}$
the position operators along the laser propagation directions determined by
the wave vectors ${\bf k}_{l}.$ Instead of 8 lasers as proposed in \cite{tam}%
, here we use only 5 lasers. The first four lasers are directed as sketched
in Fig. 1, each of which is detuned to the third lower sideband of the ion
vibrational motion, {\it i.e. }$\omega _{l=1,2,3,4}=\Delta -3\nu .$ As for
the fifth laser, its direction is unimportant but its frequency must be in
resonant with the ion transition, {\it i.e. }$\omega _{l=5}=\Delta .$
Assuming an equal wave number for all the lasers we are left with a unique
Lamb-Dicke parameter $\eta $ which is a measure of the ion's localization
region compared with the laser wavelength. If $\nu $ is much larger than any
other characteristic frequencies (resolved sideband limit) and $\eta \ll 1$
(Lamb-Dicke limit), we have, in the interaction picture and in leading order
in $\eta ,$ the interaction Hamiltonian of the form \cite{tam} 
\begin{equation}
{\cal H}_{int}=\left[ -\frac{i\eta ^{3}}{6}\sum_{l=1}^{4}\Omega _{l}\text{e}%
^{-i\phi _{l}}\widehat{A}_{l}^{3}+\Omega _{5}\text{e}^{-i\phi _{5}}\right]
\sigma _{+}+\text{H.c.}  \label{F}
\end{equation}
where the annihilation operators of the vibration quanta in the direction $%
{\bf R}_{l}$ are denoted by $\widehat{A}_{l\text{ }}$which are related to $%
\widehat{a}_{x(y,z)}$ as 
\begin{equation}
\widehat{A}_{1}=\widehat{a}_{x}+\widehat{a}_{y}+\widehat{a}_{z},\text{ }%
\widehat{A}_{2}=\widehat{a}_{x}-\widehat{a}_{y}+\widehat{a}_{z},  \label{A}
\end{equation}
\begin{equation}
\widehat{A}_{3}=\widehat{a}_{x}+\widehat{a}_{y}-\widehat{a}_{z},\text{ }%
\widehat{A}_{4}=\widehat{a}_{x}-\widehat{a}_{y}-\widehat{a}_{z}.  \label{Al}
\end{equation}
If we now adjust the laser intensity and phase to meet the conditions $%
\Omega _{l=1,2,3,4}=\Omega $ and $\phi _{l=1,3}=\phi _{l=2,4}+\pi =\phi ,$
then the Hamiltonian (\ref{F}) simplifies to 
\[
{\cal H}_{int}=\zeta \left( \widehat{a}_{x}\widehat{a}_{y}\widehat{a}%
_{z}-\xi \right) \sigma _{+}+\text{H.c.} 
\]
with $\zeta $ and $\xi $ determined by the system parameters in a
controllable manner as 
\begin{equation}
\zeta =-6i\eta ^{3}\Omega \exp [-(i\phi +\eta ^{2}/2)],  \label{zeta}
\end{equation}
\begin{equation}
\xi =-\frac{i\Omega _{5}}{6\Omega \eta ^{3}}\exp [i(\phi -\phi _{5})].
\label{xi}
\end{equation}
Because the damping of vibrational quanta is negligible, the main decay
process occurs via ionic spontaneous emission with rate $\gamma .$ Thus, in
the Lamb-Dicke limit, the system density operator $\rho $ obeys the master
equation \cite{tam}

\begin{equation}
\frac{d\rho }{dt}=-i[{\cal H}_{int},\rho ]-\frac{\gamma }{2}\left( \sigma
_{+}\sigma _{-}\rho +\rho \sigma _{+}\sigma _{-}-2\sigma _{-}\rho \sigma
_{+}\right) .  \label{rr}
\end{equation}
In the long-time limit, Eq. (\ref{rr}) has a ``dark'' steady solution $\rho
_{s}=\left| g\right\rangle \left| \Psi \right\rangle \left\langle \Psi
\right| \left\langle g\right| $ with $\left| \Psi \right\rangle $
responsible for the steady state of the ion vibration. It is easy to verify
that $\left| \Psi \right\rangle $ obeys the equation 
\begin{equation}
\widehat{a}_{x}\widehat{a}_{y}\widehat{a}_{z}\left| \Psi \right\rangle =\xi
\left| \Psi \right\rangle .  \label{a}
\end{equation}
Further, if the ion is initially prepared in a Fock state $\left| \Psi
_{0}\right\rangle =\left| q+k\right\rangle _{x}\left| p+k\right\rangle
_{y}\left| k\right\rangle _{z}$ with $k$ a non-negative integer, then,
because $p$ and $q$ are conserved in the problem under treatment, the state $%
\left| \Psi \right\rangle $ satisfies also the two equations 
\begin{equation}
\widehat{P}\left| \Psi \right\rangle =p\left| \Psi \right\rangle ,\text{ }
\label{b}
\end{equation}
\begin{equation}
\widehat{Q}\left| \Psi \right\rangle =q\left| \Psi \right\rangle .  \label{c}
\end{equation}
Comparing (\ref{a}) - (\ref{c}) with (\ref{e1}) - (\ref{e3}) yields $\left|
\Psi \right\rangle \equiv \left| \xi ,p,q\right\rangle ${\it \ }which is the
vibrational TCS we want to generate.

To follow the transient process and to assess generation time ({\it i.e. }%
the time needed for the system to reach the steady regime) we simulate Eq. (%
\ref{rr}) by the Monte Carlo Wave-Function approach \cite{mcwf}. The system
wave function at time $t$ is looked for in the form 
\begin{equation}
\Phi (t)=\sum_{l,m,n}\left[ G_{lmn}(t)\left| g\right\rangle
+E_{lmn}(t)\left| e\right\rangle \right] \left| l\right\rangle _{x}\left|
m\right\rangle _{y}\left| n\right\rangle _{z}.  \label{Phi}
\end{equation}
For the purpose of numerical calculation, the Fock-state basis should be
confined to $l,m,n\leq N_{\max }$ with $N_{\max }$ a cutoff to be chosen
such that its increasing does not change the result within a pre-set high
enough accuracy. In our simulations we have found $N_{\max }=10$ sufficient
for most of the numerical calculation: just an error of less than $10^{-6}$
arises by increasing the cutoff by one. For convenience, we use the
dimensionless time $\tau $ determined by $\tau =\gamma t.$ For the initial
condition $\Phi (0)=\left| e\right\rangle \left| q+k\right\rangle _{x}\left|
p+k\right\rangle _{y}\left| k\right\rangle _{z}$ with $k=4,$ the time
evolution of the probability $P(l,m,n)=|G_{lmn}|^{2}+|E_{lmn}|^{2}$ of
finding the ion in a Fock state $\left| l\right\rangle _{x}\left|
m\right\rangle _{y}\left| n\right\rangle _{z}$ is plotted in Fig. 2a. As is
visual from that figure, at the beginning $P(l,m,n)$ oscillate but at large $%
\tau $ they approach $|C_{n}(\xi ,p,q)|^{2}$ (see Eq. (\ref{C})) precisely,
implying generation of the vibrational TCS. The initial variation of $%
P(l,m,n)$ depends on the value of $k$ but their long-time behavior is $k$%
-independent. This means that the TCS is eventually generated independently
of the choice of $k$ which is best taken as $k=0$ because the state $\left|
e\right\rangle \left| q\right\rangle _{x}\left| p\right\rangle _{y}\left|
0\right\rangle _{z}$ is easiest to prepare practically. The transition
towards the steady regime can be watched as well in Fig. 2b showing how the
inversion $\left\langle \sigma _{z}\right\rangle =\sum_{l,m,n}\left(
|E_{lmn}|^{2}-|G_{lmn}|^{2}\right) $ evolves from $1$ to $-1.$ The purity of
the generated state is examined by the fidelity $F(\tau )=|\left\langle \Phi
(\tau )\right| \left. \xi ,p,q\right\rangle |^{2}$ whose change in time is
depicted in Fig. 3. Because of the truncation in the Fock basis we determine
the vibrational TCS generation time $\tau _{s}$ by the constraint $1-F(\tau
_{s})=10^{-5}$ and display the $\tau _{s}$ as a function of $\alpha =\zeta
/\gamma $ in Fig. 4a. The decreasing of $\tau _{s}$ with increasing $\alpha $
plays a role as will be seen.

Finally, we discuss a possibility to make the generation time in the present
improved scheme ($L=5$ scheme) shorter than that in the previous scheme ($%
L=8 $ scheme) \cite{tam}. Suppose we wish to produce a state $\left| \xi
,p,q\right\rangle $ with fixed $\xi ,$ $p$ and $q.$ The desired values of $p$
and $q$ are easily to manage by preparing an initial state in the form $%
\left| e\right\rangle \left| q+n\right\rangle _{x}\left| p+n\right\rangle
_{y}\left| n\right\rangle _{z}$ with $n$ a non-negative integer. To tailor $%
\xi $ we use Eq. (10) of Ref. \cite{tam} and Eq. (\ref{xi}) of this work.
The same value of $\xi $ in the $L=8$ scheme can be obtained in the $L=5$
scheme in either of the two following ways: (i) keeping the frequency of the
5th laser in the $L=5$ scheme equal to that of the 8th laser in the $L=8$
scheme, but decreasing the frequencies of the 1st to 4th laser in the $L=5$
scheme 6 times smaller than those of the 1st to 7th laser in the $L=8$
scheme or (ii) increasing the frequency of the 5th laser in the $L=5$ scheme
6 times larger than that of the 8th laser in the $L=8$ scheme, but keeping
the frequencies of the 1st to 4th laser in the $L=5$ scheme equal to those
of the 1st to 7th laser in the $L=8$ scheme. In case (i) the value of $\zeta 
$ in the $L=5$ scheme (see Eq. (\ref{zeta}) of this work) remains unchanged
in comparison with that in the $L=8$ scheme (see the equation before Eq.
(10) in Ref. \cite{tam}) and, hence, both schemes undergo the same dynamics.
Nevertheless, if case (ii) is chosen, the value of $\zeta $ in the $L=5$
scheme becomes 6 time larger than that in the $L=8$ scheme. Since $\tau
_{s}, $ for the same value of $\gamma ,$ decreases with increasing $\zeta $
(see Fig. 4a), a net advantage in the generation time appears in the
improved scheme. Figure 4b shows that advantage from an angle of the $\xi $%
-dependence, for more clarity. A noteworthy merit of the $L=5$ scheme is
that, for the whole range of $\xi ,$ the desired TCS is generated in a
remarkably shorter time compared with the $L=8$ scheme.

In summary, we have improved a recently proposed scheme for generating trio
coherent states in the vibrational motion of an ion trapped in three
dimensions. The trio coherent state generated by this scheme is stable
because it appears in a steady regime in which the ion has fully relaxed to
its ground ``dark'' state (Fig. 2b) which is identified by the quenched
fluorescence from the ion. Besides reduction in the number of driving lasers
to be used, the improved version can also provide a much shorter generation
time. The possibility to shorten the generation time $\tau _{s}$ relies upon
its decrease with increasing $\alpha .$ In general, such a decreasing
behavior does not exist for the whole range of $\alpha .$ In fact, $\tau
_{s} $ would turn out to increase with $\alpha ,$ say, when $\alpha $ is
very large. For instance, in the extreme limit of $\alpha \rightarrow \infty 
$ (corresponding to negligible spontaneous emission, $\gamma \rightarrow 0)$%
, periodic oscillations or phenomena like collapses and revivals, undamped
self-pulsing or even chaos would occur. However, in problems like the one
under consideration, where relaxation to a steady state within a finite time
is important, $\gamma $ is essential and cannot be treated as arbitrarily
small and, thus, arbitrarily large values of $\alpha $ are irrelevant. To
specify the range of $\alpha $ for which our simulations have been carried
out, we use $\eta $ as a measure of smallness since $\eta \ll 1$ in the
Lamb-Dicke limit we are working in. As explained, because the ionic decay
cannot be neglected, let us be interested in situations such that $\gamma $
satisfies the condition $\gamma /(6\Omega )\gg {\cal O}(\eta ^{3}).$ Then,
from Eq. (\ref{zeta}), we get $\alpha =\zeta /\gamma \simeq {\cal O}(6\Omega
\eta ^{3}/\gamma )$ which is much less than unity. For smaller values of $%
\gamma $ ({\it i.e. }larger $\alpha ),$ as mentioned above, abundant
dynamics associated with other different timescales will take place which
should be dealt with separately ({\it e.g., }by means of multiple-timescale
analyses).

\noindent {\bf Acknowledgments}

The authors thank the KIAS Quantum Information Group for useful discussions.
H.S.Y. is supported by a KAIST Grant for Basic Science, B.A.N. by a KIAS
Research Fund (No. 02-0149-001) and J.K. by a Korea Research Foundation
Grant (KRF-2002-070-C00029).

\newpage

\begin{center}
{\bf Figure captions}
\end{center}

\begin{enumerate}
\item[Fig. 1:]  The configuration of used lasers : the 1st (2nd, 3rd and
4th) laser propagates along the direction connecting the coordinate origin
``o'' with the point with $\{x,y,z\}=\{1,1,1\}$ $(\{1,-1,1\},$ $\{1,1,-1\}$
and $\{1,-1,-1\}),$ while the 5th laser can be oriented arbitrarily (not
shown).

\item[Fig. 2:]  a) Left: The probability $P(l,m,n)$ {\it versus }scaled time%
{\it \ }$\tau =\gamma t$ for the initial condition $\Phi (0)=\left|
e\right\rangle \left| q+4\right\rangle _{x}\left| p+4\right\rangle
_{y}\left| 4\right\rangle _{z}.$ The parameters used are $\alpha =0.02,$ $%
\xi =2.0,$ $p=3$ and $q=2.$ The values of $l,m,n$ are indicated by $(l,m,n).$
All the $P(l,m,n)$ with $m\neq n+p$ and $l\neq n+q$ are identically zeros,
whereas all the $P(q+n,p+n,n)$ with $n\geq 3$ are negligible in the
long-time limit. Right: Distribution of $|C_{n}(\xi ,p,q)|^{2}$ (see Eqs. (%
\ref{fs}) and (\ref{C})) for the same values of $\xi ,$ $p$ and $q$ as
above. The values of $n$ are indicated by $(n)$. All the $|C_{n}(\xi
,p,q)|^{2}$ with $n\geq 3$ are negligibly small. b) Time evolution of the
ionic inversion $\left\langle \sigma _{z}\right\rangle $ for the same
initial condition and parameters as in a). The inset enlarges the variation
for the interval $\tau \in [0,20].$

\item[Fig. 3:]  Fidelity $F$ as a function of $\tau $ for $\xi =2.0,$ $p=3,$ 
$q=2$ and $\alpha =0.01,$ $0.02,$ $0.03$ (curves indicated by ``a'', ``b'',
``c'', respectively).

\item[Fig. 4:]  a) The generation time $\tau _{s}$ as a function of $\alpha $
for $\xi =2.0,$ $p=3$ and $q=2.$ b) As in a) but in logarithmic scale and as
a function of $\xi $ in the $L=8$ and $L=5$ schemes. Here we use $\alpha
=0.02$ in the $L=8$ scheme which corresponds to $\alpha =0.12$ in the $L=5$
scheme.
\end{enumerate}

\end{document}